\pdfoutput=1
\documentclass[12pt,a4paper]{article}

\usepackage{ifthen} 
\newboolean{pdflatex}
\setboolean{pdflatex}{true} 

\newboolean{articletitles}
\setboolean{articletitles}{true} 

\newboolean{uprightparticles}
\setboolean{uprightparticles}{false} 

\def\paperauthors{LHCb collaboration} 
\def\paperasciititle{Observation of the doubly charmed baryon decay Xicc++ to Xic'+ pi+} 
\def\papertitle{Observation of the doubly charmed baryon decay ${\it\Xi_{cc}^{++} \to \Xi_c^{\prime +} \pi^+}$} 
\def\paperkeywords{{High Energy Physics}, {LHCb}} 
\def\papercopyright{\the\year\ CERN for the benefit of the LHCb collaboration} 
\def\paperlicence{CC BY 4.0 licence}
\def\paperlicenceurl{https://creativecommons.org/licenses/by/4.0/}

\usepackage[top=1in, bottom=1.25in, left=1in, right=1in]{geometry}

\usepackage{microtype}
\usepackage{lineno}  
\usepackage{xspace} 
\usepackage{caption} 

\usepackage{graphicx}  
\usepackage{color}
\usepackage{colortbl}
\graphicspath{{./figs/}} 

\usepackage{amsmath} 
\usepackage{amssymb}
\usepackage{amsfonts}
\usepackage{upgreek} 

\newcommand*\patchAmsMathEnvironmentForLineno[1]{%
\expandafter\let\csname old#1\expandafter\endcsname\csname #1\endcsname
\expandafter\let\csname oldend#1\expandafter\endcsname\csname
end#1\endcsname
 \renewenvironment{#1}%
   {\linenomath\csname old#1\endcsname}%
   {\csname oldend#1\endcsname\endlinenomath}%
}
\newcommand*\patchBothAmsMathEnvironmentsForLineno[1]{%
  \patchAmsMathEnvironmentForLineno{#1}%
  \patchAmsMathEnvironmentForLineno{#1*}%
}
\AtBeginDocument{%
\patchBothAmsMathEnvironmentsForLineno{equation}%
\patchBothAmsMathEnvironmentsForLineno{align}%
\patchBothAmsMathEnvironmentsForLineno{flalign}%
\patchBothAmsMathEnvironmentsForLineno{alignat}%
\patchBothAmsMathEnvironmentsForLineno{gather}%
\patchBothAmsMathEnvironmentsForLineno{multline}%
\patchBothAmsMathEnvironmentsForLineno{eqnarray}%
}

\usepackage{hyperxmp}

\usepackage[pdftex,
            pdfauthor={\paperauthors},
            pdftitle={\paperasciititle},
            pdfkeywords={\paperkeywords},
            pdfcopyright={Copyright (C) \papercopyright},
            pdflicenseurl={\paperlicenceurl}]{hyperref}

\usepackage[colorinlistoftodos,textsize=scriptsize]{todonotes}

\usepackage[bottom,flushmargin,hang,multiple]{footmisc}

\usepackage[all]{hypcap} 

\usepackage{xspace} 
\usepackage{upgreek}

\def\lhcb   {\mbox{LHCb}\xspace}

\def\MagUp {\mbox{\em Mag\kern -0.05em Up}\xspace}

\ifthenelse{\boolean{uprightparticles}}%
{

 \def\Ppi         {\ensuremath{\uppi}\xspace}

 \def\PDelta      {\ensuremath{\Delta}\xspace}                 
 \def\PXi         {\ensuremath{\Xi}\xspace}                 
 \def\PLambda     {\ensuremath{\Lambda}\xspace}                 
 \def\PSigma      {\ensuremath{\Sigma}\xspace}                 
 \def\POmega      {\ensuremath{\Omega}\xspace}                 
 \def\PUpsilon    {\ensuremath{\Upsilon}\xspace}

 \def\PB      {\ensuremath{\mathrm{B}}\xspace}                 
                  
 \def\PD      {\ensuremath{\mathrm{D}}\xspace}

 \def\PK      {\ensuremath{\mathrm{K}}\xspace}

 \def\Pb      {\ensuremath{\mathrm{b}}\xspace}                 
 \def\Pc      {\ensuremath{\mathrm{c}}\xspace}                 
 \def\Pd      {\ensuremath{\mathrm{d}}\xspace}

 \def\Pi      {\ensuremath{\mathrm{i}}\xspace}

 \def\Pp      {\ensuremath{\mathrm{p}}\xspace}

 \def\Ps      {\ensuremath{\mathrm{s}}\xspace}                 
                  
 \def\Pu      {\ensuremath{\mathrm{u}}\xspace}

 \def\thebaroffset{0.0em}
}
{

 \def\Ppi         {\ensuremath{\pi}\xspace}

 \mathchardef\PDelta="7101
 \mathchardef\PXi="7104
 \mathchardef\PLambda="7103
 \mathchardef\PSigma="7106
 \mathchardef\POmega="710A
 \mathchardef\PUpsilon="7107
                  
 \def\PB      {\ensuremath{B}\xspace}                 
                  
 \def\PD      {\ensuremath{D}\xspace}

 \def\PK      {\ensuremath{K}\xspace}

 \def\Pb      {\ensuremath{b}\xspace}                 
 \def\Pc      {\ensuremath{c}\xspace}                 
 \def\Pd      {\ensuremath{d}\xspace}

 \def\Pi      {\ensuremath{i}\xspace}

 \def\Pp      {\ensuremath{p}\xspace}

 \def\Ps      {\ensuremath{s}\xspace}                 
                  
 \def\Pu      {\ensuremath{u}\xspace}

 \def\thebaroffset{0.18em}
}
\newcommand{\offsetoverline}[2][\thebaroffset]{\kern #1\overline{\kern -#1 #2}}%

\makeatletter
\ifcase \@ptsize \relax
  \newcommand{\miniscule}{\@setfontsize\miniscule{4}{5}}
\or
  \newcommand{\miniscule}{\@setfontsize\miniscule{5}{6}}
\or
  \newcommand{\miniscule}{\@setfontsize\miniscule{5}{6}}
\fi
\makeatother

\DeclareRobustCommand{\optbar}[1]{\shortstack{{\miniscule (\rule[.5ex]{1.25em}{.18mm})}
  \\ [-.7ex] $#1$}}

\def\uquark    {{\ensuremath{\Pu}}\xspace}

\def\dquark    {{\ensuremath{\Pd}}\xspace}

\def\squark    {{\ensuremath{\Ps}}\xspace}

\def\cquark    {{\ensuremath{\Pc}}\xspace}

\def\bquark    {{\ensuremath{\Pb}}\xspace}

\def\pion   {{\ensuremath{\Ppi}}\xspace}
\def\piz    {{\ensuremath{\pion^0}}\xspace}
\def\pip    {{\ensuremath{\pion^+}}\xspace}
\def\pim    {{\ensuremath{\pion^-}}\xspace}

\def\kaon    {{\ensuremath{\PK}}\xspace}

\def\KorKbar {\kern \thebaroffset\optbar{\kern -\thebaroffset \PK}{}\xspace}

\def\Km      {{\ensuremath{\kaon^-}}\xspace}

\def\D       {{\ensuremath{\PD}}\xspace}

\def\DorDbar {\kern \thebaroffset\optbar{\kern -\thebaroffset \PD}\xspace}

\def\Dp      {{\ensuremath{\D^+}}\xspace}
\def\Dm      {{\ensuremath{\D^-}}\xspace}

\def\DpDm    {\ensuremath{\Dp {\kern -0.16em \Dm}}\xspace}

\def\B       {{\ensuremath{\PB}}\xspace}

\def\BorBbar {\kern \thebaroffset\optbar{\kern -\thebaroffset \PB}\xspace}

\def\Bd      {{\ensuremath{\B^0}}\xspace}

\def\BdorBdbar {\kern \thebaroffset\optbar{\kern -\thebaroffset \Bd}\xspace}

\def\Bs      {{\ensuremath{\B^0_\squark}}\xspace}

\def\BsorBsbar {\kern \thebaroffset\optbar{\kern -\thebaroffset \Bs}\xspace}

\def\Y#1S{\ensuremath{\PUpsilon{(#1S)}}\xspace}

\def\proton      {{\ensuremath{\Pp}}\xspace}

\def\Lz          {{\ensuremath{\PLambda}}\xspace}

\def\LorLbar     {\kern \thebaroffset\optbar{\kern -\thebaroffset \PLambda}\xspace}

\def\Xires       {{\ensuremath{\PXi}}\xspace}

\def\Lc          {{\ensuremath{\Lz^+_\cquark}}\xspace}

\def\Xic         {{\ensuremath{\Xires_\cquark}}\xspace}

\def\Xicp        {{\ensuremath{\Xires^+_\cquark}}\xspace}

\def\Xiccp       {{\ensuremath{\Xires^+_{\cquark\cquark}}}\xspace}
\def\Xiccpp      {{\ensuremath{\Xires^{++}_{\cquark\cquark}}}\xspace}

\def\BF         {{\ensuremath{\mathcal{B}}}\xspace}
\def\BR         {\BF}

\def\to                 {\ensuremath{\rightarrow}\xspace}

\def\AT#1     {\ensuremath{A_{\mathrm{T}}^{#1}}\xspace}           

\def\C#1      {\ensuremath{\mathcal{C}_{#1}}\xspace}                       
\def\Cp#1     {\ensuremath{\mathcal{C}_{#1}^{'}}\xspace}                    
\def\Ceff#1   {\ensuremath{\mathcal{C}_{#1}^{\mathrm{(eff)}}}\xspace}        
\def\Cpeff#1  {\ensuremath{\mathcal{C}_{#1}^{'\mathrm{(eff)}}}\xspace}       
\def\Ope#1    {\ensuremath{\mathcal{O}_{#1}}\xspace}                       
\def\Opep#1   {\ensuremath{\mathcal{O}_{#1}^{'}}\xspace}                    

\newcommand{\nospaceunit}[1]{\ensuremath{\text{#1}}}       
\newcommand{\aunit}[1]{\ensuremath{\text{\,#1}}}       

\newcommand{\tev}{\aunit{Te\kern -0.1em V}\xspace}
\newcommand{\gev}{\aunit{Ge\kern -0.1em V}\xspace}
\newcommand{\mev}{\aunit{Me\kern -0.1em V}\xspace}
\newcommand{\kev}{\aunit{ke\kern -0.1em V}\xspace}
\newcommand{\ev}{\aunit{e\kern -0.1em V}\xspace}
 
\newcommand{\mevc}{\ensuremath{\aunit{Me\kern -0.1em V\!/}c}\xspace}
\newcommand{\gevc}{\ensuremath{\aunit{Ge\kern -0.1em V\!/}c}\xspace}
\newcommand{\mevcc}{\ensuremath{\aunit{Me\kern -0.1em V\!/}c^2}\xspace}
\newcommand{\gevcc}{\ensuremath{\aunit{Ge\kern -0.1em V\!/}c^2}\xspace}

\def\mum  {\ensuremath{\,\upmu\nospaceunit{m}}\xspace}

\def\fb   {\ensuremath{\aunit{fb}}\xspace}
\def\invfb   {\ensuremath{\fb^{-1}}\xspace}

\def\fs   {\aunit{fs}}

\def\gsim{{~\raise.15em\hbox{$>$}\kern-.85em
          \lower.35em\hbox{$\sim$}~}\xspace}
\def\lsim{{~\raise.15em\hbox{$<$}\kern-.85em
          \lower.35em\hbox{$\sim$}~}\xspace}

\def\sPlot{\mbox{\em sPlot}\xspace}

\def\sqs   {\ensuremath{\protect\sqrt{s}}\xspace}

\def\pt         {\ensuremath{p_{\mathrm{T}}}\xspace}

\def\ptot       {\ensuremath{p}\xspace}

\def\evtgen     {\mbox{\textsc{EvtGen}}\xspace}

\def\geant      {\mbox{\textsc{Geant4}}\xspace}

\def\photos     {\mbox{\textsc{Photos}}\xspace}

\def\pythia     {\mbox{\textsc{Pythia}}\xspace}

\def\tell1  {TELL1\xspace}
\def\ukl1   {UKL1\xspace}

\def\Xiccpp       {{\ensuremath{\Xires^{++}_{\cquark\cquark}}}\xspace}

\def\Xiccp        {{\ensuremath{\Xires^{+}_{\cquark\cquark}}}\xspace}

\def\Xicprime     {{\ensuremath{\Xires^{'+}_\cquark}}\xspace}
\def\XicPrime     {{\ensuremath{\Xires^{'+}_\cquark}}\xspace}

\def\XiccSigDecay	{{\ensuremath{\Xiccpp\to\Xicprime\pip}}\xspace}
\def\XiccCtrlDecay	{{\ensuremath{\Xiccpp\to\Xicp\pip}}\xspace}
\def\XiccCtrDecay	{{\ensuremath{\Xiccpp\to\Xicp\pip}}\xspace}

\def\XicDecay      {{\ensuremath{\Xicp\to\proton\Km\pip}}\xspace}

\def\relativeRatioTOS	{{\ensuremath{1.81 \pm 0.43 \pm 0.25}}\xspace}
\def\relativeRatioTIS	{{\ensuremath{1.34 \pm 0.19 \pm 0.11}}\xspace}

\def\relativeRatioabstract	{{\ensuremath{1.41 \pm 0.17 \pm 0.10}}\xspace}

\usepackage{cite} 
\usepackage{mciteplus}

\usepackage{longtable} 

\begin{document}

\renewcommand{\thefootnote}{\fnsymbol{footnote}}
\setcounter{footnote}{1}


\begin{titlepage}
\pagenumbering{roman}

\vspace*{-1.5cm}
\centerline{\large EUROPEAN ORGANIZATION FOR NUCLEAR RESEARCH (CERN)}
\vspace*{1.5cm}
\noindent
\begin{tabular*}{\linewidth}{lc@{\extracolsep{\fill}}r@{\extracolsep{0pt}}}
\ifthenelse{\boolean{pdflatex}}
{\vspace*{-1.5cm}\mbox{\!\!\!\includegraphics[width=.14\textwidth]{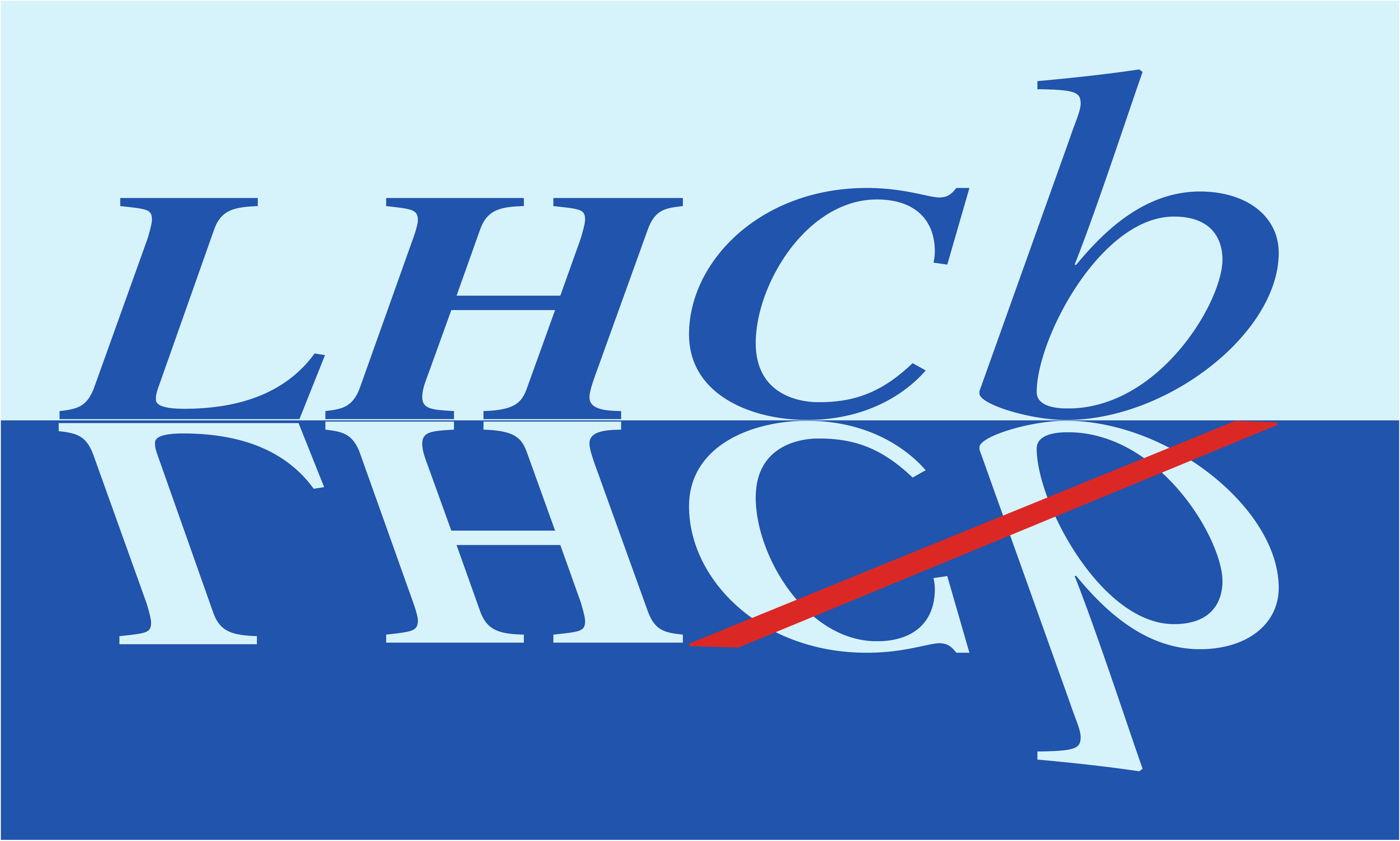}} & &}%
{\vspace*{-1.2cm}\mbox{\!\!\!\includegraphics[width=.12\textwidth]{figs/lhcb-logo.eps}} & &}%
\\
 & & CERN-EP-2022-016 \\  
 & & LHCb-PAPER-2021-052 \\  
 & & 06 May 2022 \\ 
& &  \\
\end{tabular*}

\vspace*{4.0cm}

{\normalfont\bfseries\boldmath\huge
\begin{center}
  \papertitle 
\end{center}
}

\vspace*{2.0cm}

\begin{center}
\paperauthors\footnote{Authors are listed at the end of this paper.}
\end{center}

\vspace{\fill}

\begin{abstract}
  \noindent
  The \XiccSigDecay decay is observed 
  using proton-proton collisions collected by the \lhcb experiment at a centre-of-mass energy of 13\tev,
  corresponding to an integrated luminosity of $5.4\invfb$.
  The \XiccSigDecay decay is reconstructed partially, where the photon
  from the $\PXi_c^{\prime+} \to \Xicp\gamma$ decay is not reconstructed and the $pK^-\pi^+$ final state of the $\Xicp$ baryon is employed.
  The \XiccSigDecay branching fraction relative to that of the \XiccCtrlDecay decay is measured to be
\relativeRatioabstract,
 where the first uncertainty is statistical and the second systematic. 
  
\end{abstract}

\vspace*{2.0cm}

\begin{center}
  Published in JHEP 05 (2022) 038
\end{center}

\vspace{\fill}

{\footnotesize 
\centerline{\copyright~\papercopyright. \href{\paperlicenceurl}{\paperlicence}.}}
\vspace*{2mm}

\end{titlepage}


\newpage
\setcounter{page}{2}
\mbox{~}
%
%
%
%


\renewcommand{\thefootnote}{\arabic{footnote}}
\setcounter{footnote}{0}

\cleardoublepage


\pagestyle{plain} 
\setcounter{page}{1}
\pagenumbering{arabic}


\section{Introduction}
\label{sec:Introduction}
 
The quark model~\cite{GellMann:1964nj,Zweig:352337, *Zweig:570209} predicts the existence of doubly charmed baryons that contain two charm quarks
and a light quark (\uquark, \dquark, \squark), 
providing ideal systems to test effective theories of quantum chromodynamics (QCD). 
In 2017, the LHCb collaboration reported the first observation of the doubly charmed baryon \Xiccpp using the decay into the $\Lc\Km\pip\pip$ final state, and measured its mass~\cite{LHCb-PAPER-2017-018}.\footnote{The inclusion of charge-conjugate processes is implied throughout.} 
This observation has been confirmed using the $\mbox{\XiccCtrlDecay}$ decay mode~\cite{lhcb-paper-2018-026}, as proposed by Ref.~\cite{XiccppPrediction}.
The LHCb collaboration also measured the $\Xiccpp$ lifetime~\cite{lhcb-paper-2018-019} and the production rate~\cite{lhcb-paper-2019-035}, 
and established an upper limit for the $\Xiccpp\to \Dp p \Km \pip$ decay mode~\cite{lhcb-paper-2019-011}.
The $\Xiccpp$ mass has been measured~\cite{LHCb-PAPER-2019-037} using both the $\Xiccpp\to\Lc\Km\pip\pip$ and \XiccCtrlDecay
 decays. 
 
 This paper presents the observation of 
 the \XiccSigDecay decay and the measurement of
its branching fraction relative to that of the \XiccCtrlDecay transition, 
\begin{equation}
    \frac{\BR(\XiccSigDecay)}{\BR(\XiccCtrDecay)},
\end{equation}
using proton-proton ($pp$) collisions collected by the \lhcb experiment at a centre-of-mass energy of 13\tev,
  corresponding to an integrated luminosity of $5.4\invfb$.
  The signal \mbox{\XiccSigDecay} decay is partially reconstructed, with the photon
  from the $\PXi_c^{\prime+} \to \Xicp\gamma$ process not reconstructed. 
The $\Xicp$ baryon is reconstructed with the \XicDecay decay for both the signal and normalisation modes. 
This measurement
can be used to test various theoretical models, by comparing the measured relative branching fraction to theoretical predictions. 
There is a wide spread in these predicted values, due to several theory assumptions. 

\begin{figure}[tb]
    \centering
    \includegraphics[width=0.9\textwidth]{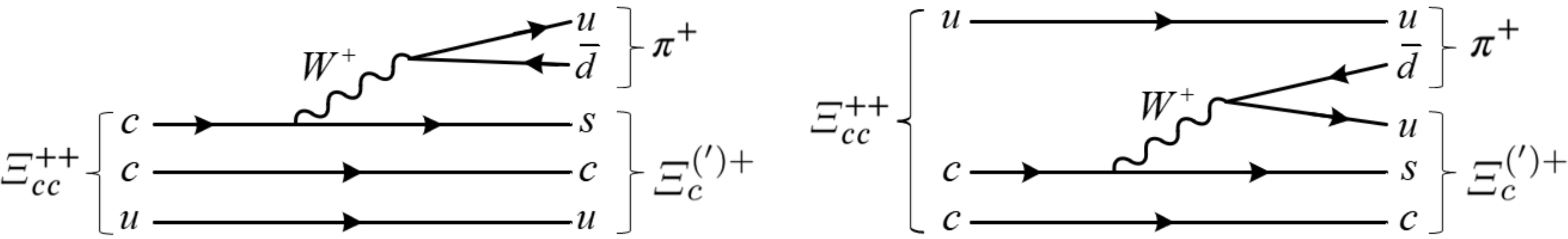}
    \caption{
    (Left) external and (right) internal $W$-emission diagrams of the \mbox{$\Xiccpp\to\Xires_{\cquark}^{(')+}\pip$} decay.
    }
    \label{Feynman_diagram}
\end{figure}

Two topological diagrams contribute to the $\Xiccpp\to\Xires_{c}^{(\prime)+}\pip$ decay amplitude,
as shown in Fig.~\ref{Feynman_diagram}, corresponding to external and internal $W$-emission.
The contribution of the internal $W$-emission can vary.
Including the contribution from the internal $W$-emission, the relative branching fraction is predicted between 0.81 and 0.83~\cite{Res:0.820.81, Res:0.83}. Additionally, including the rescattering mechanism between the final-state hadrons, the predicted relative branching fraction varies between 0.44 and 0.70~\cite{Han:2021azw}.
When the interference between the external and the internal $W$-emission contributions for both $S$- and $P$-wave amplitudes are considered,
the relative branching fraction is predicted to be considerably enhanced at 6.74~\cite{Res:6.74}. 

The flavour wave-function symmetry can affect the relative branching ratio.
The flavour wave-function of the \Xiccpp and \Xicp baryons is antisymmetric,
while it is symmetric for the \XicPrime state,
which implies that the $\Xiccpp\to\Xicp X$ transition is flavour symmetric while $\Xiccpp\to\XicPrime X$ is flavour antisymmetric.
As predicted by the Körner-Pati-Woo theorem \cite{KPW,KPW2},
the internal $W$-emission amplitude of the $\Xiccpp\to\XicPrime X$ transition is suppressed due to its flavour antisymmetry.
Including the Körner-Pati-Woo theorem, the relative branching fraction is predicted to be 4.33~\cite{Res:4.33-1,Res:4.33-2,Res:4.33-3} and 4.55~\cite{Res:4.55}.

Finally, two models are considered that approximate the internal structure and the weak-decay of the \Xiccpp state by 
treating two of the three quarks as a diquark system.
The first model assumes a (\cquark\uquark)\cquark configuration,
where the single \cquark quark decays and the diquark remains  a spectator,
leading to a prediction of the relative branching fraction of 0.70~\cite{Res:0.7}.
The second model takes the (\cquark\cquark)\uquark configuration with the diquark system breaking apart,
leading to a $(\cquark\cquark)\uquark\to\cquark(\squark\uquark)$ transition and a predicted relative branching fraction of $0.56\pm0.18$~\cite{Res:0.56} and $0.30\pm0.24$~\cite{Res:0.56, Res:0.303}.

The rest of this paper is organised as follows. In Sec.~\ref{sec:Detector}, a brief introduction to the LHCb detector and the simulation framework is given.
Sections~\ref{selection} and~\ref{BR_measurement} describe the event selection and the measurement of the \XiccSigDecay branching fraction relative to the \XiccCtrlDecay decay. The systematic uncertainties related to this measurement are reported in Sec.~\ref{SystematicUncertainties}. Finally, the results are summarized in Sec.~\ref{summary}.


\section{Detector and simulation}
\label{sec:Detector}

The \lhcb detector~\cite{LHCb-DP-2008-001,LHCb-DP-2014-002} is a single-arm forward
spectrometer covering the \mbox{pseudorapidity} range $2<\eta <5$,
designed for the study of particles containing \bquark or \cquark
quarks. The detector includes a high-precision tracking system
consisting of a silicon-strip vertex detector surrounding the $pp$
interaction region, a large-area silicon-strip detector located
upstream of a dipole magnet with a bending power of about
$4{\mathrm{\,Tm}}$, 
and three stations of silicon-strip detectors and straw
drift tubes placed downstream of the magnet.
The tracking system provides a measurement of the momentum, \ptot, of charged particles with
a relative uncertainty that varies from 0.5\% at low momentum to 1.0\% at 200\gevc.
The minimum distance of a track to a primary \mbox{$pp$ collision} vertex, the impact parameter, 
is measured with a resolution of $(15+29/\pt)\mum$,
where \pt is the component of the momentum transverse to the beam, in\,\gevc.
Different types of charged hadrons are distinguished using information
from two ring-imaging Cherenkov detectors. 
Photons, electrons and hadrons are identified by a calorimeter system consisting of
scintillating-pad and preshower detectors, an electromagnetic
and a hadronic calorimeter. 
Muons are identified by a system composed of alternating layers of iron and multiwire proportional chambers.

The online event selection is performed by a trigger~\cite{LHCb-DP-2012-004}, 
which consists of a hardware stage, based on information from the calorimeter and muon
systems, followed by two software stages, which apply partial and full event reconstructions sequentially.
At the first software stage one or two tracks with a large impact parameter significance and $\pt$ is required. 
Then an alignment and calibration of the detector is performed in near real-time~\cite{LHCb-PROC-2015-011}.
This process allows the reconstruction of \Xiccpp decays to be performed entirely in the second stage of the software trigger~\cite{LHCb-DP-2016-001}, whose output is used as input to the present analysis.

Simulation samples are used to model the effects of the detector acceptance and to estimate the efficiencies of the selection requirements.
In the simulation, \mbox{$pp$ collisions} are generated using
\pythia 8~\cite{Sjostrand:2007gs,*Sjostrand:2006za} 
with a specific \lhcb configuration~\cite{LHCb-PROC-2010-056}.
A dedicated generator \textsc{GenXicc}2.0~\cite{Chang:2009va} is used for the $\Xiccpp$ production.
Decays of unstable particles are described by \evtgen~\cite{Lange:2001uf}, in which final-state radiation is generated using \photos~\cite{davidson2015photos}. 
The interaction of the generated particles with the detector, and its response, are implemented using the \geant toolkit~\cite{Allison:2006ve}.
Fast simulated samples generated with the \textsc{AmpGen}~\cite{AmpGen,*AmpGen2} and \textsc{RapidSim}~\cite{RapidSim} toolkits are also used to study the signal distribution with different amplitude hypotheses and to estimate various background decays that may appear in the data,
such as $\Xires_{\cquark\cquark}^{++}\to\Xires_\cquark(2645/2790)^{+}\pip$ decays.
Simulated events are generated with a $\Xiccpp$ mass of $3621\mevcc$~\cite{LHCb-PAPER-2019-037} and a lifetime of $256\fs$~\cite{lhcb-paper-2018-019}.
 
\section{Event selection}
\label{selection}
The reconstruction and event selection of the $\Xiccpp\to\Xires_{\cquark}^{(')+}\pip$ decays are the same as in the previous \Xiccpp \lhcb analysis~\cite{LHCb-PAPER-2019-037}
except for the trigger requirements.
Two positively and one negatively charged tracks, corresponding to the final-state particles of the $\Xicp\to pK^-\pi^+$ candidate, are required to form a good quality vertex. An additional, positively charged track  is 
combined with the $\Xicp$ candidate to form a $\Xiccpp\to\Xires_{\cquark}^{(')+}\pip$ candidate with a second good quality vertex. The two vertices and all tracks are required to be detached from any primary \mbox{$pp$ collision} vertex.
Particle identification (PID) is required on the four tracks.
A multilayer perceptron (MLP) algorithm from the \textsc{Tmva} toolkit~\cite{Hocker:2007ht,*TMVA4} is used to improve the signal purity.
The MLP algorithm is trained using simulated \XiccCtrlDecay events as signal proxy and wrong-sign 
$\Xicp\pim$ combinations in data as background proxy. 
Variables associated with the \Xiccpp candidates and their decay products are used in the training.
The threshold applied to the MLP response is determined by maximising the signal significance $S/\sqrt{S+B}$, 
where $S$ and $B$ are the expected yields of signal and background in the signal region of the $\XiccCtrlDecay$ decay, respectively. 
This MLP working point also works well for the \XiccSigDecay decay.

Compared with the previous analysis \cite{LHCb-PAPER-2019-037},  
the data samples are further split into two disjoint subsamples using information from the hardware trigger. In this way the hardware trigger efficiencies for these two subsamples are well defined.
The first contains candidates that are triggered by at least one of the $\Xicp$ decay products with high transverse energy deposited in the calorimeters, and is referred to as triggered on signal (TOS). 
The second consists of events that are exclusively triggered by particles unrelated to the signal decay products; these events can be triggered, for example, by the decay products of charmed hadrons produced together with the signal baryon, and are referred to as exclusively triggered independently of signal (TIS).

\section{Relative branching fraction measurement}
\label{BR_measurement}

To measure the branching fraction of the signal decay relative to that of the normalisation channel, both the relative signal yields and
efficiencies must be determined, as defined below,
\begin{equation}
\frac{\BR(\XiccSigDecay)}{\BR(\XiccCtrDecay)}=\frac{N_{\Xicprime}}{N_{\Xicp}}\times\frac{{\epsilon}_{\Xicp}}{{\epsilon}_{\Xicprime}},
\end{equation}
where $N_{\Xires^{(')+}_\cquark}$ is the signal yield of the $\Xiccpp\to\Xires^{(')+}_\cquark\pip$ decay, and ${\epsilon}_{\Xires^{(')+}_\cquark}$ is the total efficiency for each decay.
The relative signal yield is determined by fitting the $\Xicp\pip$ invariant-mass spectrum in the data, 
and the relative efficiency is determined from fully simulated samples of the signal and normalisation decay modes. 

\begin{figure}[tb]
\centering
\includegraphics[width=0.95\linewidth]{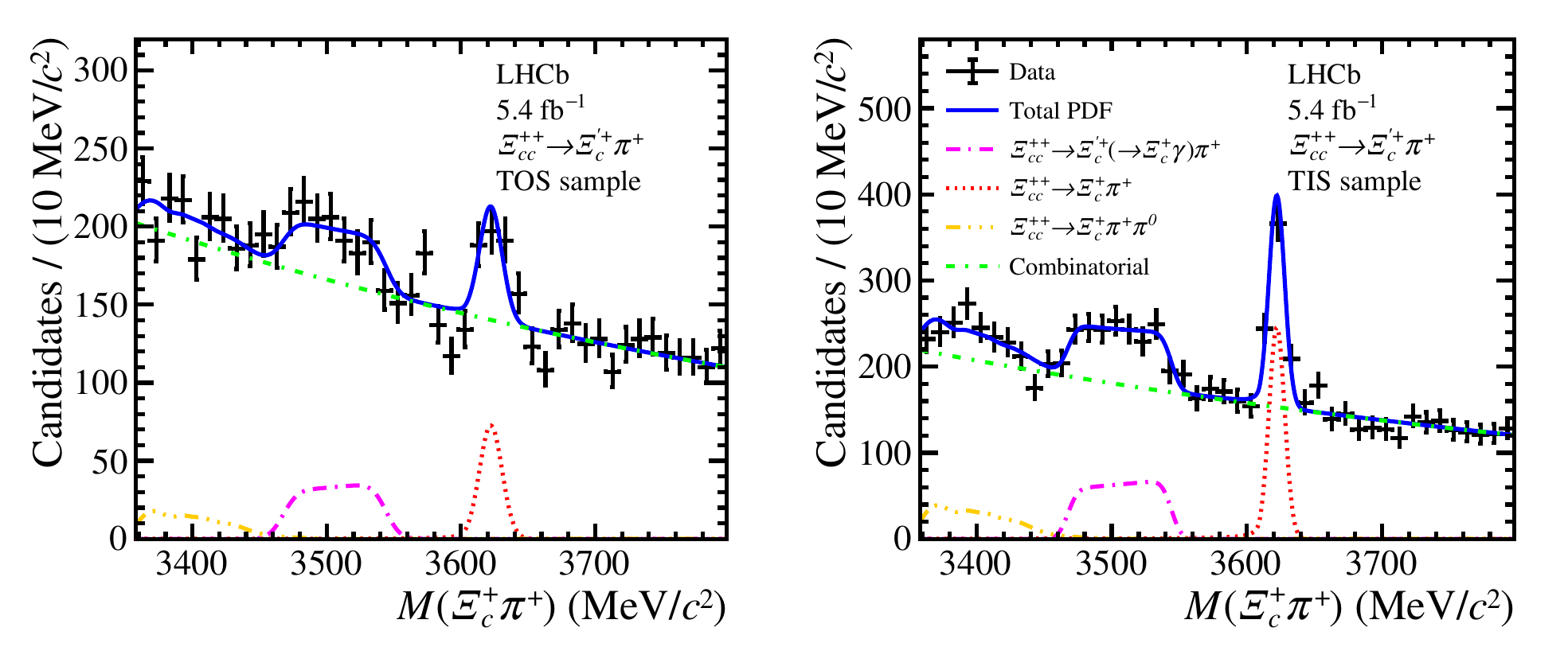}
\caption{Invariant-mass distribution of the $\Xiccpp$ candidates from the (left) TOS and (right) TIS samples, with the results of the fit overlaid. The \XiccSigDecay component is shown as a purple dashed line, the \XiccCtrlDecay component as a red dotted line, the $\Xiccpp\to\Xicp\pip\piz$ component as a yellow dashed line and the combinatorial component as a green dashed line.}
\label{fig:FitResult}
\end{figure}

The $\Xicp\pip$ invariant-mass spectrum, separated for TOS and TIS samples, are shown in Fig.~\ref{fig:FitResult}.
The peaking structure around 3620\mevcc is due to the \XiccCtrlDecay decay, and the box-like enhancement between 3480 and 3560 \mevcc is due to the \XiccSigDecay decay, shifted down and distorted because of the unreconstructed photon.

An unbinned maximum-likelihood fit is performed simultaneously to the invariant-mass
distribution 
$M(\Xicp\pip)\equiv m(\Xicp\pip)-m(\Xicp)+m_0(\Xicp)$. Here, $m(\Xicp\pip)$ and $m(\Xicp)$ are the reconstructed invariant masses of the 
$\Xiccpp$ and $\Xicp$ candidates, and $m_0(\Xicp)$ is the known \Xicp mass~\cite{PDG2020}.

Four components are considered in the fit model, separately for the TOS and TIS categories.
The \XiccCtrDecay decay is described by a Crystal Ball (CB) function~\cite{Skwarnicki:1986xj}, defined as 
\begin{equation}\label{eq:crystall-ball-function}
f(x| \alpha, n, \bar{x}, \sigma)=
\begin{cases}
 e^{-\frac{(x-\bar{x})^2}{2\sigma^2}} &{\rm for\,} \frac{x-\bar{x}}{\sigma}>-\alpha\\
 \left(\frac{n}{\alpha}\right)^n e^{-\frac{\alpha^2}{2}} \left(\frac{n}{\alpha}-\alpha-\frac{x-\bar{x}}{\sigma}\right)^{-n}  &{\rm for\,} \frac{x-\bar{x}}{\sigma}\leqslant-\alpha ,
\end{cases}
\end{equation}
where $\bar{x}$ is the mean mass and is shared between the TOS and TIS samples and $\sigma$ is the mass resolution and is varied independently in the two subsamples. The parameters $\alpha$ and $n$ describe the tail caused by the final-state radiation, and 
are parameterised as a function of the mass resolution as done in Ref.~\cite{LHCb-PAPER-2019-037}. 
The \XiccSigDecay decay is described by a limited linear function describing the true mass distribution of the signal convoluted with the Gaussian mass resolution $\sigma$. 
The limited linear function is defined as
\begin{equation}
    f\left(x| k, x_{\min }, x_{\max }\right)=\left\{\begin{array}{ll}
k \frac{x-x_{\min }}{x_{\max }-x_{\min }}-\frac{k-1}{2} & \text { for } x_{\min } \leqslant x \leqslant x_{\max } \\
0 & \text { other, }
\end{array}\right.
\end{equation}
where the lower and upper boundaries ($x_{\min}, x_{\max}$) of the function are fixed by the allowed kinematic range, while
the slope $k$ is determined from simulation. 
The distribution of the partially reconstructed background $\Xiccpp\to\Xicp\pip\pi^0$ is taken from simulation,
where the relative yield between this component and the $\Xiccpp\to\Xicp\pip$ decay 
in the TOS sample is fixed according to that in the TIS sample and the relative efficiencies between these two decays in the simulated TIS and TOS samples. The combinatorial background in each sample is described by an exponential function, with their slopes allowed to vary freely in the fit.

The invariant-mass distribution of the $\Xiccpp$ candidates together with the fit results for the two trigger categories are shown in Fig.~\ref{fig:FitResult}.
The signal and normalisation yields determined from the fit, along with the relative yields, are listed in Table~\ref{tab:yields},
where the quoted uncertainties are statistical only. The statistical significance of the \XiccSigDecay decay is 9.6 standard deviations, obtained by applying a likelihood ratio test to fits with and without the signal component.

\begin{table}[t]
\caption{Yields of the signal and normalisation decay modes, and the relative yields.} 
\centering
\begin{tabular}{cccc}
 Category & \XiccSigDecay & \XiccCtrlDecay & ${N_{\Xicprime}}/{N_{\Xicp}}$ \\\hline
 TOS & $262\pm53$ & $159\pm32$ & $1.64\pm0.39$\\
 TIS & $494\pm63$ & $379\pm32$ & $1.30\pm0.18$\\
\end{tabular}
\label{tab:yields}
\end{table}

Fully simulated samples of the signal and the normalisation modes are used to evaluate the relative efficiencies.
For both modes, the kinematic distributions in simulation, including 
the $\Xiccpp$
transverse momentum and event multiplicity,
are weighted to match those in data,
separately for the TOS and TIS categories.
The background-subtracted distributions in the data are determined with the \sPlot method~\cite{splot}, using $M(\Xicp\pip)$ as discriminating variable
in the region between $3570$ and $3680$\mevcc,
where only the $\XiccCtrlDecay$ decay and combinatorial background are present.
The fit model for these components is the same as described previously.
The relative efficiencies are determined to be
$1.105\pm0.050$ and $1.029\pm0.037$ for TOS and TIS samples, respectively, where
the uncertainties are statistical.

\section{Systematic uncertainties}
\label{SystematicUncertainties}
The relative branching fraction measurement has systematic uncertainties arising from determinations of the relative signal yields and efficiencies, as summarised in Table~\ref{tab:SysErr}.

\begin{table}
  \caption{Relative systematic uncertainties on the branching fraction ratio $\frac{\BR (\XiccSigDecay)}{\BR(\XiccCtrDecay)}$.} 
  \centering
      \begin{tabular}{l r@{.}l c}
Source	& \multicolumn{2}{l}{TOS [\%]} & TIS [\%]\\ \hline
Signal model  &  4&9  &  0.8 \\
normalisation model  &  3&7  &  3.8 \\
Combinatorial background  &  0&6  &  3.1 \\
Partially reconstructed background  &  3&7  &  1.5 \\
Mass window  &  11&0  &  3.9 \\ 

Simulated sample size  &  4&5  &  3.6 \\
Lifetime and kinematic corrections  &  0&5  &  1.8 \\
Hardware trigger &0&0 & 1.6 \\
Particle identification  &  0&5  &  0.7 \\ \hline
Sum in quadrature  &  \ \ 13&9  &  7.9 \\ 
      \end{tabular}
\label{tab:SysErr}
\end{table}

Uncertainties from the determination of the relative signal yields are caused by imperfect modeling of each component in the invariant-mass fit.
To estimate such effects, alternative models are used, replacing the corresponding components discussed in Sec.~\ref{BR_measurement}, and the changes of the relative signal yields are taken as systematic uncertainties.
A template from the fully simulated signal sample is used as an alternative signal, and replaces the limited linear function convoluted with the resolution function, leading to a systematic uncertainty of 4.9\% (0.8\%) for the TOS (TIS) sample. The alternative model for the normalisation decay mode is a CB function with the tail parameters $\alpha$ and $n$ varied in the fit, leading to a systematic uncertainty of 3.7\% (3.8\%). For the combinatorial background, a second-order polynomial function is used, and a systematic uncertainty of 0.6\% (3.1\%) is assigned.

To estimate the uncertainties from partially reconstructed background, the following decay channels are generated and reconstructed with the same final-state particles as the default, using the fast simulation toolkits~\cite{AmpGen,*AmpGen2,RapidSim}, assuming different angular momentum hypotheses ($S$-, $P$-, $D$-wave or a mixture if possible), whereas the
additional neutral pion(s), the negatively charged pion and the photon are unreconstructed:
\begin{itemize}
\item $\Xiccpp\to\Xicp\rho^+,\ \rho^+\to\pip\piz$, phase-space and different angular momentum hypotheses;

\item $\Xiccpp\to\Xicp\pip\piz$ without intermediate states;

\item
$\Xiccpp\to\Xic(2645)^+\pip,\ \Xic(2645)^+\to\Xicp\piz$, with different angular momentum hypotheses;

\item
$\Xiccp\to\Xic(2645)^0\pip, \ \Xic(2645)^0\to\Xicp\pim$;

\item
$\Xiccpp\to\Xic(2790)^+\pip, \ \Xic(2790)^+\to\Xicprime\piz, \ \PXi_c^{\prime+} \to \Xicp\gamma$;

\item
$\Xiccpp\to\Xic(2815)^+\pip, \ \Xic(2815)^+\to\Xic(2645)^+\piz,\ \Xic(2645)^+\to\Xicp\piz$.
\end{itemize}

Including the first two sources, the fits give similar results as the default fit. The statistical significance of the third and fourth decay modes is less than two standard deviations. The $M(\Xicp\pip)$ distributions of the last two decay modes are outside the mass window used in this analysis and can be ignored. Including all sources of the partially reconstructed background, the largest deviation from the default relative signal yields is taken as a systematic uncertainty of 3.7\% (1.5\%).

As the signal decay is partially reconstructed, and there is no low-mass sideband to constrain the background shape in the signal region, uncertainties due to the range of the chosen invariant-mass window are also evaluated. The $M(\Xicp\pip)$ upper sideband is primarily combinatorial background, its shape is well described by the default parameterization, therefore,
only the effects from the low-mass window boundary are considered.  
The low-mass window boundary is varied from 3350 to 3450\mevcc, and the largest deviation from the default relative signal yield is taken as a systematic uncertainty, leading to 11.0\% (3.9\%).
Although this is the largest contribution among all the systematic uncertainties, the relative branching fraction is still dominated by the statistical uncertainty.

Three sources of systematic uncertainty arising from the determination of the relative efficiencies are evaluated. First, the uncertainty due to the limited size of the simulated samples contributes 4.5\% (3.6\%).
The second is due to the $\Xiccpp$ lifetime and kinematic corrections to the simulation.
The $\Xiccpp$ lifetime~\cite{lhcb-paper-2018-019} is varied within its uncertainty, and the resulting change of the relative efficiency is taken as a systematic uncertainty. 
For the kinematic corrections, different binning schemes are used to determine the weights, which are varied by one standard deviation. Thus, the uncertainties from both the weighting method and the limited size of the background-subtracted data are considered. The relative uncertainty due to the $\Xiccpp$ lifetime and kinematic corrections to the simulation,
is less than 1\% for the TOS and 1.8\% for the TIS samples.
There is an additional photon in the \XiccSigDecay decay compared to the normalisation decay, which can pass the photon-related hardware trigger in the TIS category. By excluding such a contribution, the relative efficiency of the TIS category changes by 
1.6\%, which is taken as a systematic uncertainty. 
The last contribution arises from uncertainty on the PID efficiency and is studied using calibration samples~\cite{LHCb-DP-2018-001}. The relative uncertainty is found to be less than 1\% for both the TOS and TIS samples.
As the signal and normalisation decay modes have the same final states and very similar kinematic distributions, other systematic sources, such as the tracking efficiency,  mostly cancel in the ratio and are found to be negligible. 

\section{Results and summary}
\label{summary}
Including all systematic uncertainties, the measured relative branching fraction in the TOS and TIS samples are \relativeRatioTOS and \relativeRatioTIS, respectively,
where the first uncertainty is statistical and the second systematic.
The combination of the two measurements is performed using the best linear unbiased estimator \cite{Lyons:1988rp,Valassi:2003mu,Nisius:2020jmf,Nisius:2014wua}. 
In the combination, uncertainties arising from the modelling of the signal and normalisation modes, combinatorial and partially reconstructed background, and PID efficiency
are assumed to be 100\% correlated 
between the TOS and TIS samples, while the remaining uncertainties are taken to be uncorrelated.
The combined result is
\begin{equation}
\frac{\BR(\XiccSigDecay)}{\BR(\XiccCtrDecay)}=\relativeRatioabstract. \notag
\label{eq:BRfinal}
\end{equation}

In summary, a new decay mode of the doubly charmed baryon 
\XiccSigDecay is observed in a data sample of \mbox{$pp$ collisions} collected by the LHCb experiment at a centre-of-mass energy of $\sqs=13\tev$, corresponding to an integrated luminosity of $5.4\invfb$.
This is the third observed decay mode of the $\Xiccpp$ baryon following the 
$\Xiccpp\to\Lc\Km\pip\pip$~\cite{LHCb-PAPER-2017-018} and \XiccCtrDecay~\cite{lhcb-paper-2018-026} decays.
The relative branching fraction between the \XiccSigDecay and \XiccCtrDecay decays is measured for the fist time. The result is not consistent with current
theoretical predictions~\cite{Res:0.7, Res:0.56, Res:0.303, Res:0.820.81, Res:0.83, Res:4.33-1,Res:4.33-2,Res:4.33-3,Res:4.55,Res:6.74}, and will provide inputs for future calculations. 


\section*{Acknowledgements}
%
%
\noindent We express our gratitude to our colleagues in the CERN
accelerator departments for the excellent performance of the LHC. We
thank the technical and administrative staff at the LHCb
institutes.
We acknowledge support from CERN and from the national agencies:
CAPES, CNPq, FAPERJ and FINEP (Brazil); 
MOST and NSFC (China); 
CNRS/IN2P3 (France); 
BMBF, DFG and MPG (Germany); 
INFN (Italy); 
NWO (Netherlands); 
MNiSW and NCN (Poland); 
MEN/IFA (Romania); 
MSHE (Russia); 
MICINN (Spain); 
SNSF and SER (Switzerland); 
NASU (Ukraine); 
STFC (United Kingdom); 
DOE NP and NSF (USA).
We acknowledge the computing resources that are provided by CERN, IN2P3
(France), KIT and DESY (Germany), INFN (Italy), SURF (Netherlands),
PIC (Spain), GridPP (United Kingdom), RRCKI and Yandex
LLC (Russia), CSCS (Switzerland), IFIN-HH (Romania), CBPF (Brazil),
PL-GRID (Poland) and NERSC (USA).
We are indebted to the communities behind the multiple open-source
software packages on which we depend.
Individual groups or members have received support from
ARC and ARDC (Australia);
AvH Foundation (Germany);
EPLANET, Marie Sk\l{}odowska-Curie Actions and ERC (European Union);
A*MIDEX, ANR, IPhU and Labex P2IO, and R\'{e}gion Auvergne-Rh\^{o}ne-Alpes (France);
Key Research Program of Frontier Sciences of CAS, CAS PIFI, CAS CCEPP, 
Fundamental Research Funds for the Central Universities, 
and Sci. \& Tech. Program of Guangzhou (China);
RFBR, RSF and Yandex LLC (Russia);
GVA, XuntaGal and GENCAT (Spain);
the Leverhulme Trust, the Royal Society
 and UKRI (United Kingdom).




\clearpage
\addcontentsline{toc}{section}{References}
\bibliographystyle{LHCb}
\bibliography{main,standard,LHCb-PAPER,LHCb-CONF,LHCb-DP,LHCb-TDR}

\newpage
\centerline
{\large\bf LHCb collaboration}
\begin
{flushleft}
\small
R.~Aaij$^{32}$,
A.S.W.~Abdelmotteleb$^{56}$,
C.~Abell{\'a}n~Beteta$^{50}$,
F.~Abudin{\'e}n$^{56}$,
T.~Ackernley$^{60}$,
B.~Adeva$^{46}$,
M.~Adinolfi$^{54}$,
H.~Afsharnia$^{9}$,
C.~Agapopoulou$^{13}$,
C.A.~Aidala$^{87}$
% ...

{\footnotesize \it

% ...
$^{9}$Universit{\'e} Clermont Auvergne, CNRS/IN2P3, LPC, Clermont-Ferrand, France\\
% ...
$^{13}$LPNHE, Sorbonne Universit{\'e}, Paris Diderot Sorbonne Paris Cit{\'e}, CNRS/IN2P3, Paris, France\\
% ...
$^{32}$Nikhef National Institute for Subatomic Physics, Amsterdam, Netherlands\\
% ...
$^{46}$Instituto Galego de F{\'\i}sica de Altas Enerx{\'\i}as (IGFAE), Universidade de Santiago de Compostela, Santiago de Compostela, Spain\\
% ...
$^{50}$Physik-Institut, Universit{\"a}t Z{\"u}rich, Z{\"u}rich, Switzerland\\
% ...
$^{54}$H.H. Wills Physics Laboratory, University of Bristol, Bristol, United Kingdom\\
% ...
$^{56}$Department of Physics, University of Warwick, Coventry, United Kingdom\\
% ...
$^{60}$Oliver Lodge Laboratory, University of Liverpool, Liverpool, United Kingdom\\
% ...
$^{68}$Syracuse University, Syracuse, NY, United States\\
% ...
$^{87}$University of Michigan, Ann Arbor, United States, associated to $^{68}$\\
\bigskip
% ...
}
\end{flushleft}

\end{document}